# Three-phonon mixing as a source of light-induced chirality

Y. Zhu[1], A. Vanderhaegen[1], Z. Zeng[1,2], M. Först[1], M. Fechner[1], C. Putzke[1], P. J. W. Moll[1], D. Prabhakaran[2], P. Radaelli[2], A. Cavalleri[1,2]

[1]Max Planck Institute for the Structure and Dynamics of Matter, Hamburg, Germany

[2]Department of Physics, Clarendon Laboratory, University of Oxford, Oxford, United Kingdom

**Mid- and far-infrared optical pulses have found wide applicability in experiments in which collective modes of solids are driven nonlinearly, as a means to manipulate functional properties of materials on ultrafast time scales. In the illustrative case of boron phosphate ($BPO_4$), mid-infrared excitation has been used to induce chirality in an otherwise achiral crystal. The pump frequency dependence of photo-induced chirality reveals that, while on-resonance excitation is understood in terms of the well-documented phonon rectification from two-phonon mixing, additional pathways become relevant when the pump pulse is detuned from resonance. We find that a three-mode mixing mechanism becomes important in this case, whereas the contributions from other impulsive Raman terms are negligible. The results reported here provide a quantitative foundation for nonlinear phononic lineshapes in transparent materials, and open up new opportunities for lattice control.**

## Introduction

Nonlinear phononics extends the concept of strain engineering [1–4] to ultrafast timescales by using resonantly driven optical phonons, bypassing the speed and amplitude limitations of more traditional approaches. A widely invoked mechanism for nonlinear phononics involves rectified displacements of certain modes through lattice anharmonicities [5–17]. This approach has been used to achieve the control of ferroelectricity [8,13,14], ferroaxial order [17], magnetism [11,18], and metal-to-insulator transitions [9,10,16].

When a mid-infrared (MIR) optical drive $E(t)$ is resonant with an infrared-active lattice vibrational mode $Q_1$, it drives coherent oscillations of this mode about its equilibrium position (Fig. 1a,b). At sufficiently high field strength, the driven mode can couple anharmonically to a second mode $Q_2$, where the lowest-order nonlinear contribution to the lattice potential is of the form $U_{ph} = \alpha Q_1^2 Q_2$, with $\alpha$ being the coupling coefficient. This interaction leads to a shift of the parabolic potential of the coupled mode along its coordinate (Fig. 1c) [5–12]. Although the time-averaged displacement of the directly driven mode coordinate $Q_1$ remains zero, the motion induces a directional force proportional to $\langle Q_1^2 \rangle$, on the $Q_2$ coordinate (Fig. 1b), resulting in the rectified displacement of $Q_2$ (Fig. 1d).

In the case of boron phosphate ($BPO_4$, crystal structure shown in Fig. 2a [19]), linearly polarized MIR excitation was shown to transiently induce chirality by resonantly driving either of the two 19 THz E-symmetry phonon modes, $Q_{E,a}$ and $Q_{E,b}$ [6]. These excitations displace the B-symmetry modes that controls chirality, based on the nonlinear phonon-phonon coupling terms $Q_{E,a}^2 Q_B$ and

$Q_{E,b}^2 Q_B$ with a total potential

$$U_{ph} = \alpha(Q_{E,a}^2 - Q_{E,b}^2)Q_B \tag{1}$$

The rectified displacements of the B-symmetry modes in $BPO_4$ lift the degeneracy between local structures of opposite handedness, establishing a transient chiral state that can be detected through measurements of the time-resolved optical activity (Fig. 2b). Notably, the handedness of the induced chirality is controlled by the polarization of the driving field (Fig. 2c, insets) [20]. This behavior is evidenced by the sign reversal of the time-resolved optical activity between *a*- and *b*-axis excitation (Fig. 2c), in agreement with the prediction from nonlinear two-phonon interaction in Eq. (1) [6,20].

**Excitation Frequency Dependence**

In Ref. [6], the light-induced chirality was studied with resonant excitation of only one of the four pairs of E-symmetry modes, the one at 19 THz. The same mechanism is expected to apply to other phonon modes of the same E-symmetry. Due to their larger mode effective charges, the higher-frequency E-symmetry phonon modes at 28 and 33 THz are expected to yield a stronger light-induced chiral response. Furthermore, when exciting off resonance, more than one mode may contribute simultaneously, a mechanism that has not been studied in detail to date. Thirdly, additional non-resonant Raman contributions stimulated directly by the electric field of the MIR excitation pulses, are also expected, on symmetry grounds, to influence the response.

Here, narrowband MIR excitation pulses, obtained from difference frequency generation between chirped near-infrared pulses, were used to study the excitation frequency dependence of the light-induced chirality in $BPO_4$ [21,22]. The central frequencies of the excitation pulses were tuned from 18 THz to 34 THz with bandwidth of approximately 1.5 THz. For each of the excitation frequencies, the rotation of a near-infrared probe pulse under both *a*- and *b*-axis excitation was measured as a function of the probe polarization, allowing for the extraction of the average polarization rotation signal resulting from the chirality-induced optical activity [6]. The photosusceptibility at each excitation frequency was then defined as the slope of the linear pump fluence dependence of the induced rotary power (see Supplementary Materials for details of the data processing procedures).

Figure 3b shows the frequency dependent photosusceptibility over the range of three E-symmetry phonons, E(2), E(3), and E(4), identified by peaks in the frequency dependent extinction coefficient (Fig. 3a). Based on the $Q_E^2 Q_B$ two-phonon interaction, the response is expected to follow the spectral dependence of the extinction coefficient (Fig. 3a) [5–12]. However, with the high frequency resolution provided by the narrowband MIR excitation, we observe well-resolved deviations, which is especially true in the frequency ranges between different E-symmetry phonons (from 20 to 24 THz, and around 30 THz), suggesting the relevance of additional nonlinear coupling pathways beyond the $Q_E^2 Q_B$ mechanism.

To address this discrepancy, we consider three types of corrections to the $Q_E^2 Q_B$ two-phonon interaction (Fig. 4a, top panel). First, as suggested by G. Khalsa et al. [23], we consider two additional interactions involving the electric field $E$ of the MIR excitation pulse: the electro-phononic term $U_{E-ph} = \delta Q_E E Q_B$ describing the displacement of the $Q_B$ coordinate by mixing the MIR electric field $E$ with the driven mode $Q_E$ [23], and the impulsive stimulated Raman contribution $U_{ISRS} = \zeta E^2 Q_B$ which has been well studied with near-infrared and visible excitations [24,25]. Additionally, we consider the three-phonon coupling term of the form $U_{mix} =$ $\kappa_{ij} Q_{E,i} Q_{E,j} Q_B$, which involves the nonlinear interaction between two non-degenerate E-symmetry phonons $Q_{E,i}$ and $Q_{E,j}$, and one B-symmetry phonon $Q_B$ [7,10]. Taken together, the nonlinear coupling potential is

$$U = \sum_i \alpha_i \big(Q_{E,i,a}^2 - Q_{E,i,b}^2\big) Q_B + \sum_i \delta_i \big(Q_{E,i,a} E_a - Q_{E,i,b} E_b\big) Q_B + \zeta (E_a^2 - E_b^2) Q_B + \sum_{i<j} \kappa_{ij} \big(Q_{E,i,a} Q_{E,j,a} - Q_{E,i,b} Q_{E,j,b}\big) Q_B \tag{2}$$

with coupling coefficients $\alpha_i, \delta_i, \zeta, \kappa_{ij}$.

The first term $Q_{E,i}^2 Q_B$ has been the focus of previous studies on nonlinear phononics [5,7–12]. In the case of light-induced chirality in $BPO_4$, the coherent E-symmetry phonon oscillations drive a rectified displacement of the B-symmetry modes, which is significant when the pump frequency is resonant with one of the E-symmetry phonons and drives them to large amplitudes $A_{Q_{E,i}}$ (Fig. 4b, cyan curves).

The second term $Q_{E,i} E Q_B$, corresponding to the infrared resonant Raman term in [23], also yields an effective force on the B-symmetry phonon. An important characteristic of this mixing mechanism is its excitation frequency dependence. Since the B-symmetry phonon displacement $Q_B$ is proportional to $\langle Q_{E,i} E \rangle \sim A_{Q_{E,i}} A_{MIR} \cos(\varphi_{MIR} - \varphi_{Q_{E,i}})$, the effect depends not only on the oscillation amplitudes $A_{Q_{E,i}}$ of the driven modes $Q_{E,i}$ (Fig. 4b, cyan curves) and amplitude $A_{MIR}$ of the pump electric field $E$ (Fig. 4b, red curves), but also on their relative phase difference $\varphi_{MIR} - \varphi_{Q_{E,i}}$ (Fig. 4b, blue curves). The phase difference is close to zero for excitation frequencies below the phonon resonance, increases to $\pi/2$ at resonance, and approaches $\pi$ as the pump frequency is further increased. A similar coupling mechanism was previously reported by E. Mashkovich et al. [26], for which the interaction was mediated by a magnon instead of a phonon.

The third term $E^2 Q_B$, corresponding to the electronic Raman term in [23], is typically important at frequencies far from all phonons [24,25,27–29]. It induces a B-symmetry mode displacement proportional to $\langle E^2 \rangle$, and is not expected to have a strong excitation frequency dependence.

The fourth term $Q_{E,i} Q_{E,j} Q_B$ is the extension of lattice anharmonicity to the interaction of three phonons of different eigenfrequencies [10]. This term yields a displacement of the B-symmetry mode $Q_B$ proportional to $\langle Q_{E,i} Q_{E,j} \rangle \sim A_{Q_{E,i}} A_{Q_{E,j}} \cos(\varphi_{Q_{E,i}} - \varphi_{Q_{E,j}})$. Because the intrinsic frequencies of the two driven E-modes are different, their oscillations under MIR excitation are determined by the external electric field, with their frequencies set by the drive frequency for off resonance excitation. Nevertheless, they induce a displacive response similar to the effect from the $Q_{E,i}^2 Q_B$ mechanism [7,30]. Comparing the phase difference $\varphi_{MIR} - \varphi_{Q_{E,i}}$ between the drive electric field and the different $Q_{E,i}$ modes (Fig. 4b, blue curves), the phase differences $\varphi_{Q_{E,i}} - \varphi_{Q_{E,j}}$ between the two driven $Q_{E,i}$ modes is close to zero in the frequency ranges below the first phonon $E(i)$ and above the second phonon $E(j)$, and close to $-\pi$ for excitation frequencies in between the two phonons. Therefore, the direction of the force on the B-symmetry phonon flips sign as the excitation frequency crosses $E(i)$, and then flips again at $E(j)$.

**Simulations**

Time-and-space-dependent simulations of the phonon polariton propagation and nonlinear mode mixing inside the material were performed by means of time-domain numerical approaches combined with ab-initio calculations of the phonon coupling coefficients (see Supplementary Materials for details). The incident pulses were modeled as Fourier transform limited Gaussian waveforms with durations of 600 fs and varying central frequencies, consistent with the optical excitation used in our experiments. The induced geometric chirality depends linearly on the B-modes and quadratically on the E-modes [6]. The associated optical activity was calculated as

$$\rho = \sum_i g_{B,i} Q_B + \sum_i g_{E,ii}\left(Q_{E,i,a}^2 - Q_{E,i,b}^2\right) + \sum_{i<j} g_{E,ij}\left(Q_{E,i,a} Q_{E,j,a} - Q_{E,i,b} Q_{E,j,b}\right) \quad (3)$$

where the gyration coefficients $g$ were also obtained from first-principles calculations (see Supplementary Materials for details).

The frequency-dependent photosusceptibilities for all the four coupling mechanisms are shown in Fig. 4c. The contributions from the pure phononic couplings, namely the two-phonon interaction $Q_{E,i}^2 Q_B$ and the three-phonon interaction $Q_{E,i} Q_{E,j} Q_B$, are of similar size, because they correspond to lattice anharmonicities of the same order. The optical activity directly from the oscillating E-modes, $g_{E,ii} Q_{E,i}^2$ and $g_{E,ij} Q_{E,i} Q_{E,j}$ (Fig. 4c top panel, dashed lines), has a similar frequency dependence and is of similar size as the pure phononic terms (Fig. 4c top panel, solid lines). At the same time, the contribution from the infrared resonant Raman term is at least one order of magnitude smaller, and electronic Raman contribution is even smaller (Fig. 4c bottom panel).

We used the results of these simulations to fit the experimental data, considering only the terms with frequency dependence proportional to $Q_{E,i}^2$ and $Q_{E,i} Q_{E,j}$, in this way including the contributions from both the displaced B-modes as well as the directly driven E-modes, but neglecting the two Raman terms that involve direct coupling to the drive electric field. The initial parameters were taken from the first-principle calculations as described in the Supplementary Materials. Figure 5 shows the best fit incorporating both terms (black solid line), revealing good agreement with the experimental results (black dots). The individual contributions proportional to $Q_{E,i}^2$ and $Q_{E,i} Q_{E,j}$ are shown as green and olive curves, respectively.

The model considering only the $Q_{E,i}^2$ contributions cannot fully explain the measurement results, because they will show enhancement only at resonance. The introduction of additional $Q_{E,i} Q_{E,j}$ contributions is crucial to resolve this discrepancy, because their specific frequency dependence shifts the maxima in photosusceptibility away from the exact phonon resonances, in agreement with the experimental observations.

In summary, we used narrowband MIR excitations to map the excitation frequency dependence of light-induced chirality in $BPO_4$ in the MIR spectra range. The observed frequency dependence deviates from the behavior expected when only the nonlinear coupling between one driven phonon and the symmetry-reducing B-modes, $Q_{E,i}^2 Q_B$, is considered. We identified additional nonlinear coupling terms, and found that three-phonon mixing coupling of the form $Q_{E,i} Q_{E,j} Q_B$ yields a significant contribution to the observed signal. Note that although the Raman-type coupling terms of the form $Q_{E,i} E Q_B$ and $E^2 Q_B$ do not play an important role in the light-induced chirality of $BPO_4$, they might become important in other material systems.

Using spatiotemporal polariton propagation simulations based on ab-initio coupling parameters, we evaluated the relative contributions of different coupling mechanisms as a function of excitation frequency. At phonon resonances, the light-induced chirality effect is dominated by the coupling of the symmetry-reducing B-modes to individual E-symmetry modes, $Q_{E,i}^2 Q_B$, whereas at frequencies a few phonon linewidths away from the resonance of E-symmetry phonons, the mixing between multiple E-symmetry phonons, $Q_{E,i} Q_{E,j} Q_B$, becomes dominant. These findings open up new possibilities in manipulating material properties with light. Specifically, owing to its bipolar nature, the $Q_{E,i} Q_{E,j} Q_B$ term introduces the possibility to control the sign of light-matter coupling by tuning the frequency of the excitation pulses.

In the present case, the three-phonon interaction involves two driven modes with the same symmetry. A natural extension would be to investigate trilinear couplings involving three modes of distinct symmetries, $Q_X Q_Y Q_Z$, where $X$, $Y$, and $Z$ denote different irreducible representations. In such a scenario, simultaneous excitation of both $Q_X$ and $Q_Y$ could induce a displacement of $Q_Z$ exclusively through the trilinear interaction. Importantly, couplings such as $Q_X^2 Q_Z$ and $Q_Y^2 Q_Z$ would be symmetry forbidden, making the trilinear pathway the only allowed mechanism for driving the third mode [31].

The most promising avenue that one could envisage involves pairs of narrowband pump sources resonant with either of the two driven modes, $Q_X$ and $Q_Y$, resulting in an oscillatory excitation of the third mode. This effect would amount to a non-degenerate three-phonon mixing, which may lead to a parametric amplification of $Q_Z$ modes at finite frequencies [32]. It will be interesting to explore the potential to extend the trilinear phononic coupling mechanisms to other systems controllable with nonlinear phononics, including perovskites [16,33,34], magnetic systems [26,35–37], and ferroelectric materials [8,13,38].

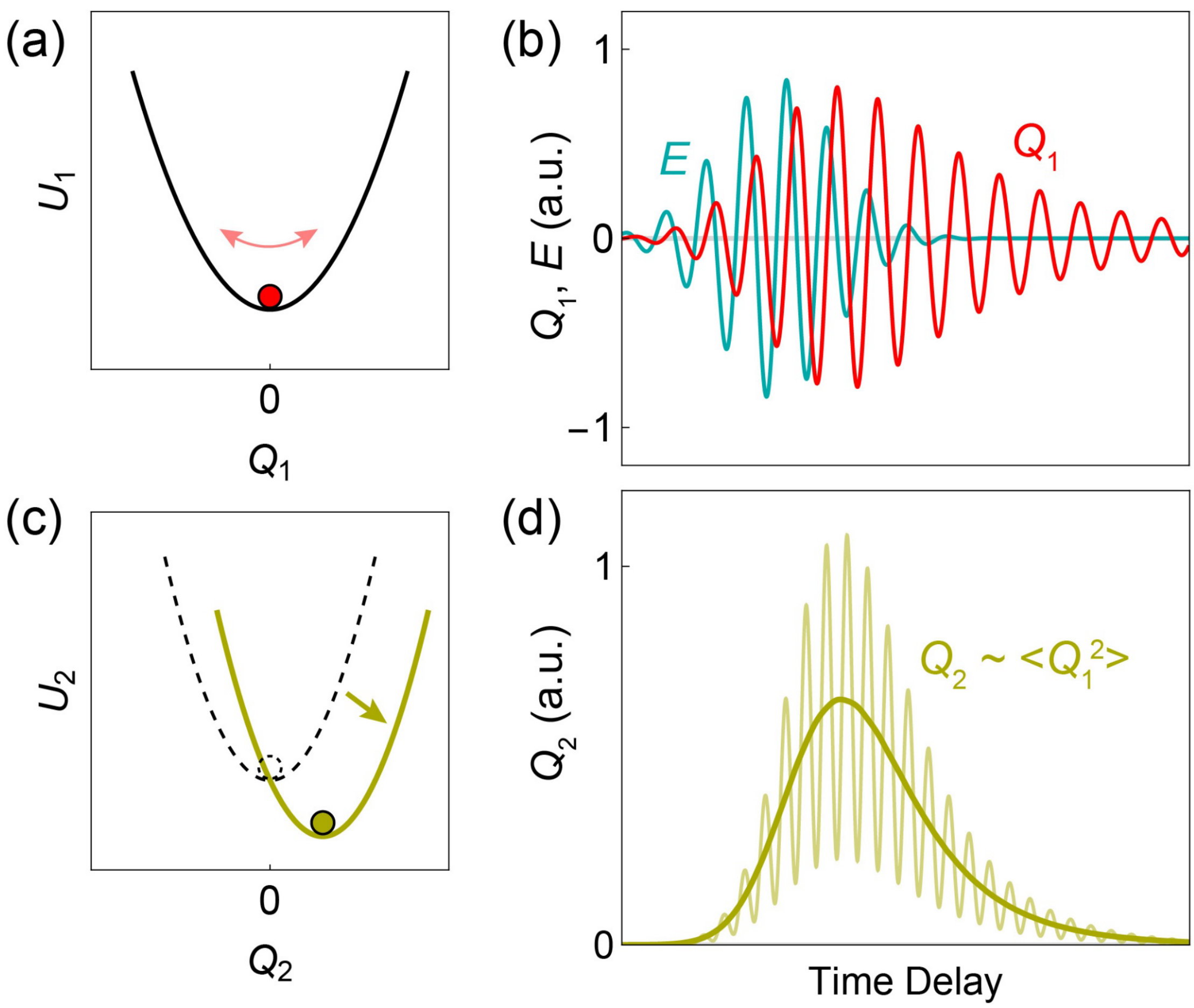

**Figure 1: The $Q_1^2 Q_2$ coupling in nonlinear phononics. (a)** Harmonic potential of an infrared-active mode $Q_1$, which couples linearly to the electric field. **(b)** Oscillatory motion of this mode $Q_1$ (red) induced by a resonant electric field drive $E$ (cyan). **(c)** Potential of mode $Q_2$ in absence (dashed) and presence (olive) of an oscillating $Q_1$ mode. The nonlinear coupling of the form $Q_1^2 Q_2$ shifts the potential minimum on average to a finite value. **(d)** Motion of mode $Q_2$ (light olive) induced by the $Q_1^2 Q_2$ coupling, together with its rectified component (dark olive).

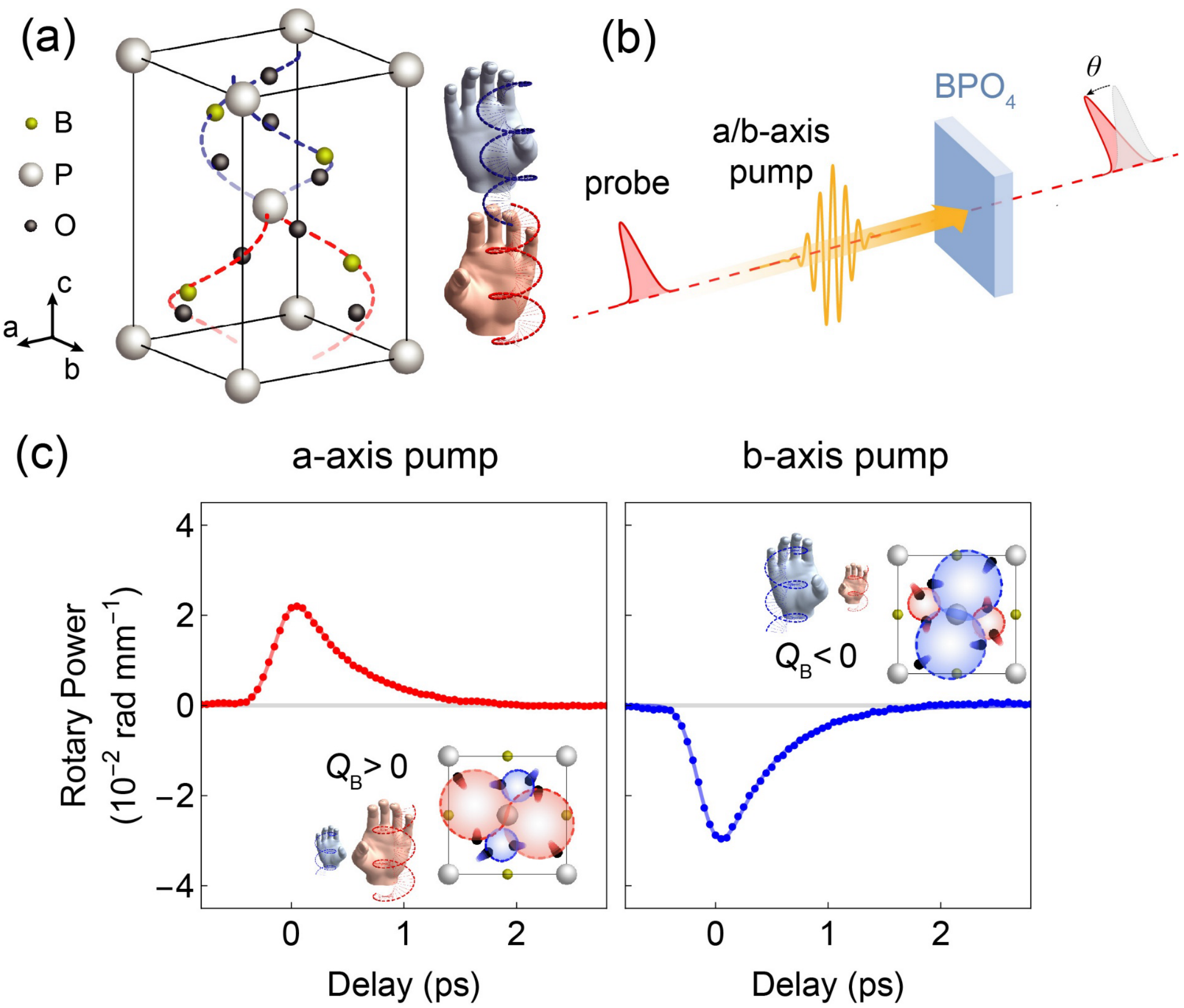

**Figure 2: Light-induced chirality in $BPO_4$. (a)** Crystal structure of $BPO_4$, composed of chiral parts with opposite handedness (red and blue) within the unit cell. **(b)** Schematic drawing of the pump probe experiment. The mid-infrared excitation pulse (orange) is polarized along either *a*- or *b*-axis. A time-delayed near-infrared pulse is used to measure the induced chirality through polarization rotation. **(c)** Time delay dependent polarization rotation, as measured in Ref. [6]. Transient chiral states, induced by *a*- and *b*-axis excitation, exhibit opposite handedness.

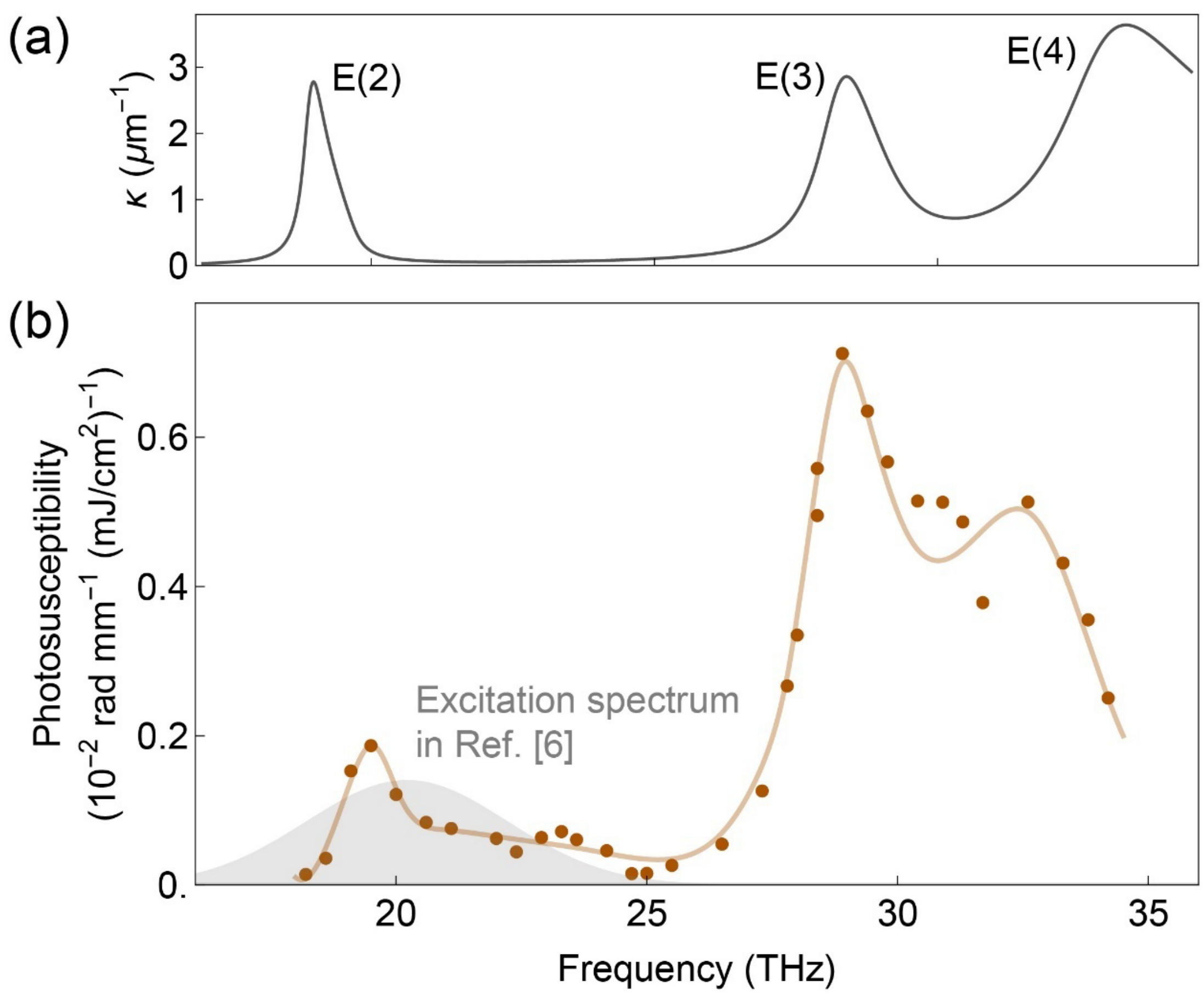

**Figure 3: Excitation frequency dependence of light-induced chirality in $BPO_4$. (a)** Extinction coefficient of $BPO_4$ in the mid-infrared spectral range for light polarized in the *ab*-plane, extracted from Fourier-transform infrared spectroscopy. Peaks correspond to three E-symmetry phonon modes. **(b)** Photosusceptibility of the induced chirality as a function of excitation frequency (dots), together with a guide to the eye. The broadband excitation spectrum from an earlier study [6] is shown as gray shade.

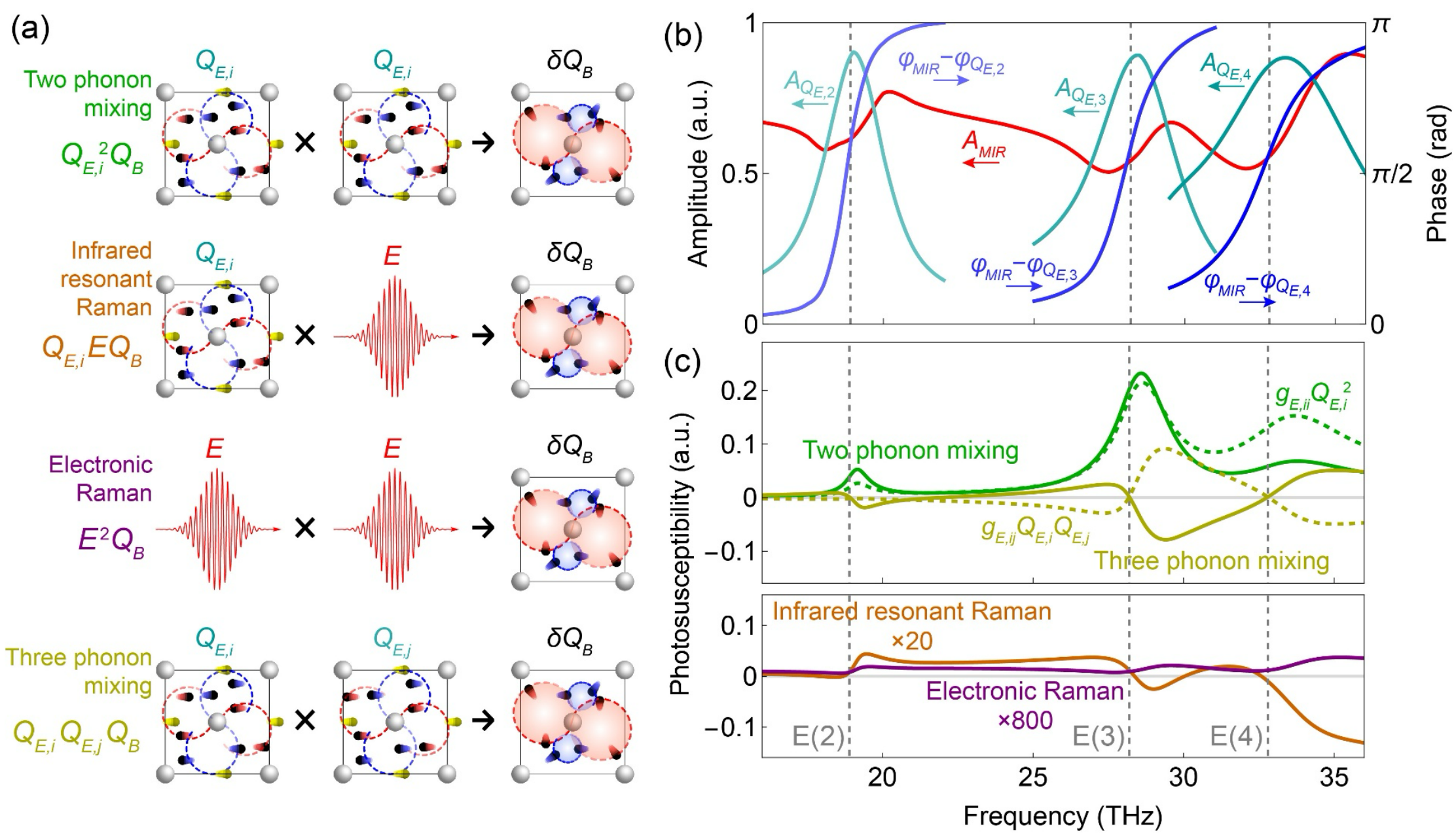

**Figure 4: Nonlinear coupling to the symmetry-reducing B-symmetry modes in $BPO_4$. (a)** Illustration of two-phonon mixing $Q_{E,i}^2 Q_B$, infrared resonant Raman scattering $Q_{E,i}EQ_B$, electronic Raman scattering $E^2Q_B$, and three-phonon mixing $Q_{E,i}Q_{E,j}Q_B$ coupling. Oscillations of the E-modes $Q_{E,i}$ and/or the electric field $E$ induce the displacement of B-symmetry modes $Q_B$. **(b)** Excitation frequency dependence of electric field oscillation amplitude $A_{MIR}$ (red), E-mode oscillation amplitude $A_{Q_E}$ (cyan), and their relative phase difference $\varphi_{MIR} - \varphi_{Q_E}$ (blue) beneath the sample surface. Vertical dashed lines indicate the TO frequencies of the E-symmetry modes. **(c)** Excitation frequency dependence of the chirality-induced optical activity, arising from B-mode displacements due to two-phonon mixing (green solid line), infrared resonant Raman scattering (orange solid line), electronic Raman scattering (purple solid line), and three-phonon mixing (olive solid line). Optical activity is also induced by the driven E-modes and proportional to terms $g_{E,ii}Q_{E,i}^2$ (green dashed line) and $g_{E,ij}Q_{E,i}Q_{E,j}$ (olive dashed line). Data are calculated from a first-principles approach, exhibiting significantly smaller responses for the Raman scattering mechanisms that involve the drive electric field directly.

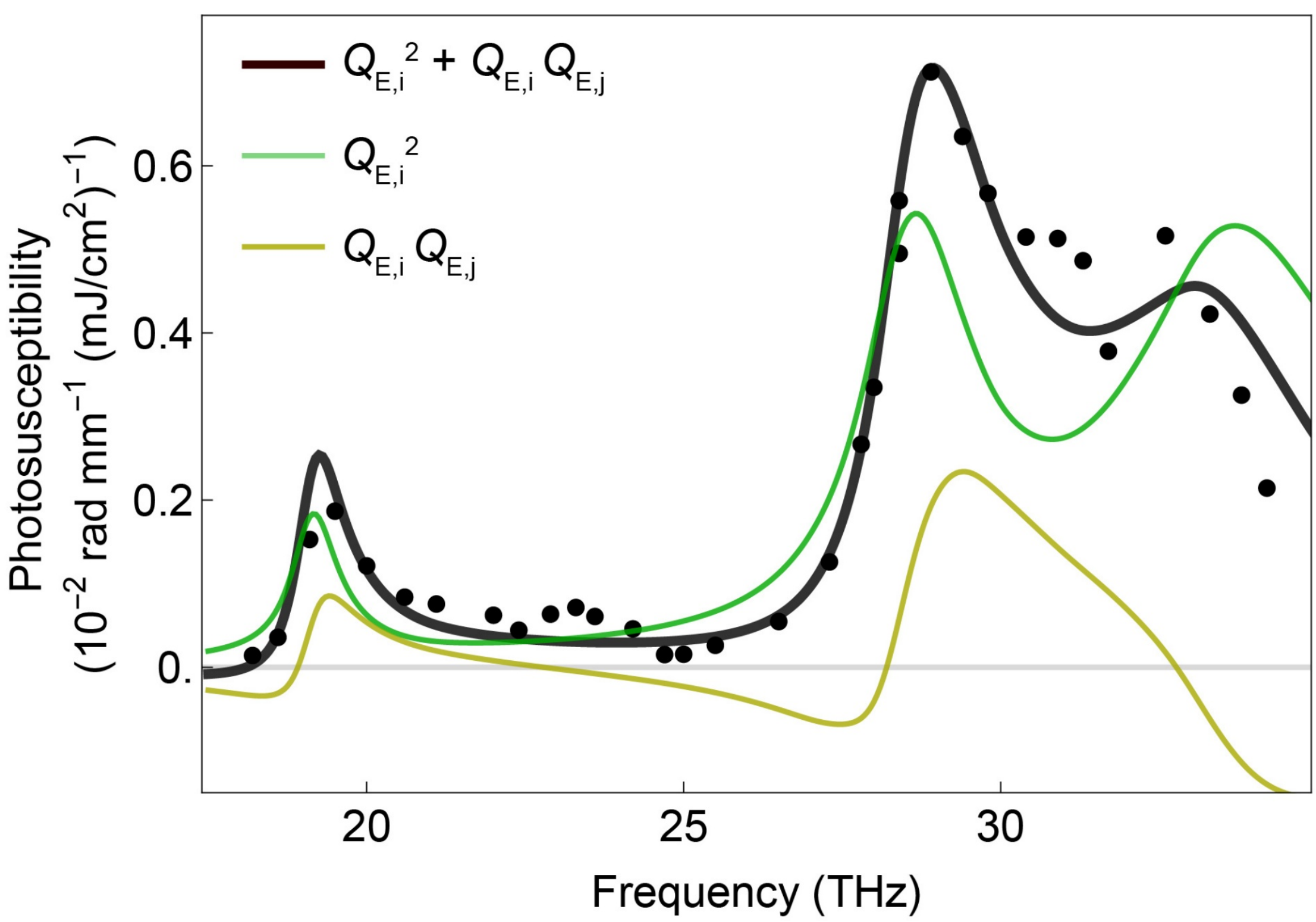

**Figure 5: Decomposition of the photosusceptibility into contributions from different coupling mechanisms.** Experimental data (black dots) are shown together with the best fit (black solid line) according to Eqs. (2) and (3), which is composed of contributions proportional to $Q_{E,i}^2$ and $Q_{E,i}Q_{E,j}$, shown as green and olive solid lines, respectively.

24956 (2024).

Supplementary Materials for

# Three-phonon mixing as a source of light-induced chirality

Y. Zhu[1], A. Vanderhaegen[1], Z. Zeng[1,2], M. Först[1], M. Fechner[1], C. Putzke[1], P. J. W. Moll[1], D. Prabhakaran[2], P. Radaelli[2], A. Cavalleri[1,2]

[1]Max Planck Institute for the Structure and Dynamics of Matter, Hamburg, Germany

[2]Department of Physics, Clarendon Laboratory, University of Oxford, Oxford, United Kingdom

**Table of Contents**

## Section I: Sample preparation and characterization

Polycrystalline $BPO_4$ powder was synthesized with high-purity $H_3BO_3$ (99.9995%) and $NH_4H_2PO_4$ (99.998%) powders, before being ground, pelletized, and sintered. Single crystals were grown from this powder using $Li_2MoO_4$ flux via the top-seeded solution growth technique in a platinum crucible at up to 975 °C. After homogenization, the melt was slowly cooled down, resulting in the growth of high-quality $BPO_4$ single crystals [1].

These crystals were characterized by X-ray diffractometry and oriented with an optically flat (001) surface using focused ion beam milling. Xenon ions were accelerated at 30 kV at a beam current of 2.5 mA for the surface orientation. To reduce roughness, the sample was finally milled at grazing incidence with a beam current of 200 nA. The thickness of the sample used in this experiment was ~200 μm.

Static terahertz (THz) to mid-infrared (MIR) optical properties of $BPO_4$ were characterized by Fourier-transform infrared reflectivity (FTIR) measurements, with light polarized along the *a*-axis [2]. The experimental result is shown in Fig. S1a, which clearly shows four E-symmetry phonon modes. They were fitted with the dielectric function

$$\tilde{\varepsilon}(\omega) = \varepsilon_\infty + \varepsilon_\infty \sum_j \frac{\omega_{LO,j}^2 - \omega_{TO,j}^2}{\omega_{TO,j}^2 - \omega^2 - i\gamma_j\omega} \qquad \text{(S1)}$$

which yielded the frequency-dependent reflectivity and extinction coefficient shown in Figs. S1b and S1c, respectively. The fitting parameters are listed in Table S1.

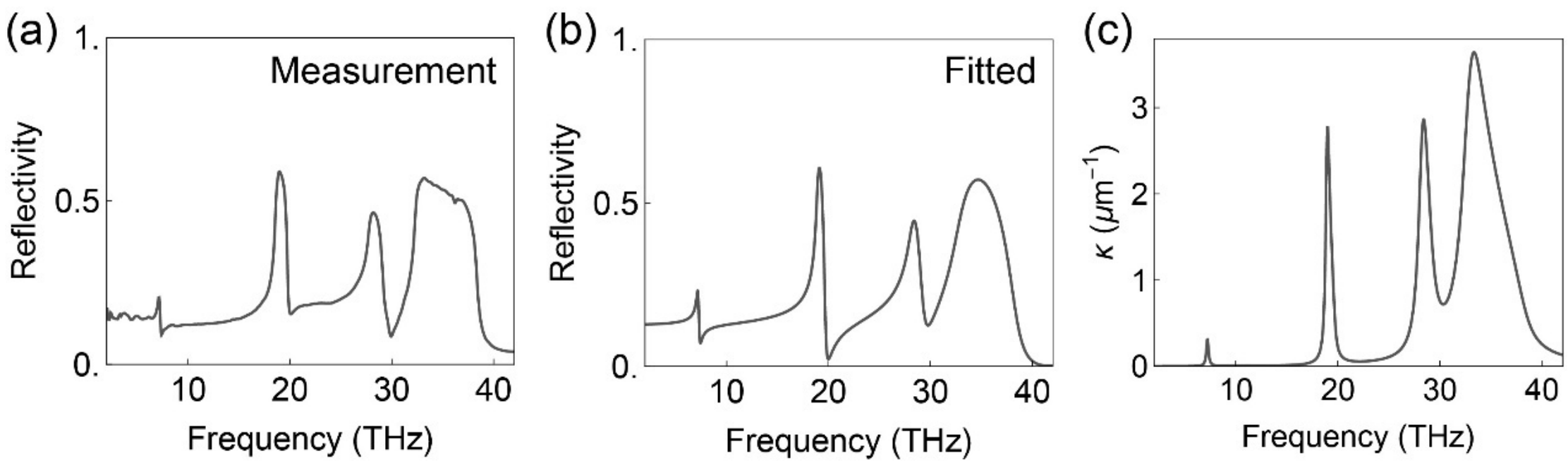

**Figure S1: Equilibrium optical properties of $BPO_4$ in the THz to MIR frequency range. (a)** Static reflectivity spectrum of $BPO_4$ for electric fields polarized along the *a*-axis. **(b)** Fit of the measured spectrum with four Lorentzian oscillators, as defined in Eq. (S1). **(c)** Extinction coefficient of $BPO_4$ extracted from the fit.

**Table S1: Fit results of four E-symmetry phonon modes in $BPO_4$ for *a*- or *b*-axis polarization.**

| Phonon Mode | $\omega_{TO,j}/2\pi$ (THz) | $\omega_{LO,j}/2\pi$ (THz) | $\gamma_j/2\pi$ (THz) | $\varepsilon_\infty$ |
|---|---|---|---|---|
| **E(1)** | 7.2 | 7.4 | 0.24 | / |
| **E(2)** | 18.9 | 20.3 | 0.33 | / |
| **E(3)** | 28.2 | 30.5 | 0.96 | / |
| **E(4)** | 32.8 | 37.0 | 2.0 | / |
| / | / | / | / | 2.70 |

**Section II: Experimental setup for time-resolved polarization rotation measurements**

The experimental setup is sketched in Fig. S2. Narrow bandwidth and frequency tunability of our MIR excitation source was achieved by chirped-pulse difference frequency generation, as described in detailed in Ref. [3]. A commercial Ti:sapphire amplifier system with 1 kHz repetition rate, ~75 fs pulse duration, ~6.5 mJ pulse energy, and 800 nm central wavelength was used to drive two identical three stage optical parametric amplifiers (OPAs), seeded with the same white light. The two wavelength-tunable near-infrared (NIR) output pulses were linearly chirped with a pair of grating-based stretchers, enabling a continuous tuning of their pulse duration. The chirped pulses were mixed in a GaSe crystal, generating linearly-polarized, narrowband MIR excitation pulses tunable between 17 and 38 THz. Typical excitation spectra of with ~1.5 THz bandwidth, measured using an FTIR setup, are shown in Fig. S3. The excitation pulses were focused onto the sample with an off-axis parabolic mirror, with a wavelength-dependent spot size of ~100 to 140 µm at sample position.

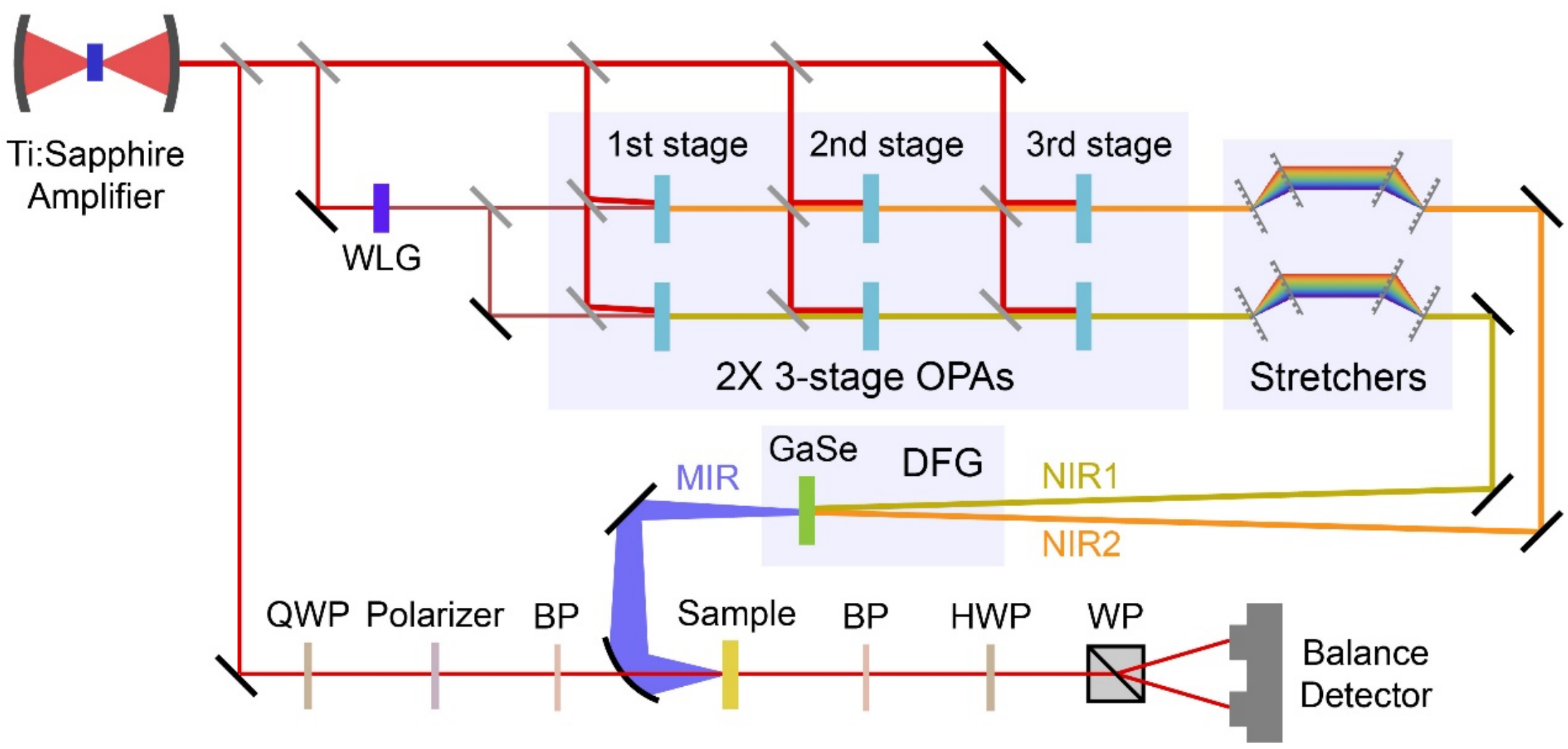

**Figure S2: Schematic drawing of the experimental setup.** WLG: white light generation, OPA: optical parametric amplifier, DFG: difference frequency generation, QWP: quarter waveplate, HWP: half waveplate, BP: bandpass filter (808/5 nm), WP: Wollaston prism.

The NIR probe pulses for polarization rotation measurements were split from the Ti:sapphire amplifier, and were focused onto the sample by a lens, reaching a spot size of <50 µm. The probe polarization was controlled by a motorized polarizer behind a quarter waveplate. The pump-induced polarization rotation was measured using a half waveplate, a Wollaston prism, and two balanced photodiodes. Two identical bandpass filters with center wavelength 808 nm and bandwidth 5 nm were placed in the probe path before and after the sample, which filter out the side bands from the Pockels effect. The pump and probe pulses were both at normal incidence to the (001) sample surface. The MIR excitation pulses were linearly polarized. The sample was rotated about the surface normal to allow for excitation along the *a*- or *b*-axis of the crystal. All experiments were performed at room temperature.

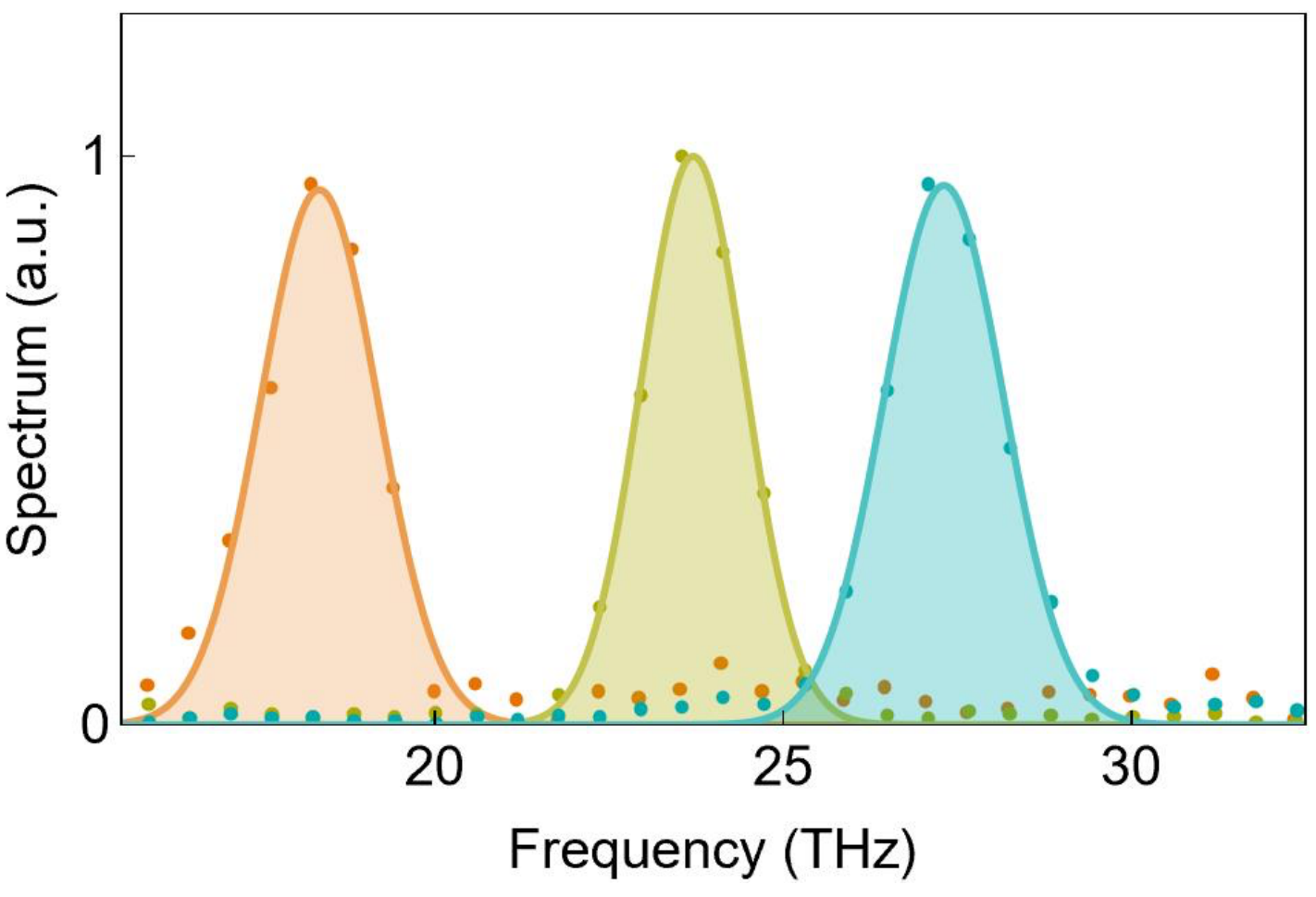

**Figure S3: Three example spectra of the MIR excitation pulses.**

**Section III: Pump and probe polarization dependence of the polarization rotation signal**

For pump polarizations along both the *a*- and *b*-axis, the time-resolved polarization rotation was measured as a function of probe incident polarization. Figures S4a,c show the full probe polarization dependence for the example excitation frequency of 33.3 THz, measured at an excitation fluence of 11 mJ/cm$^2$. The measured polarization rotation signal $\theta$ follows the angular dependence [2]

$$\theta(\varphi) = A_1\rho + A_2(\Delta n)^2\cos(4\varphi - \gamma_0) \tag{S2}$$

Here, $\varphi$ is the angle between the pump and the probe polarization, $\gamma_0$ is the angle between the pump polarization and the optical axis of photo-induced birefringence, $\rho$ is the rotary power proportional to the photo-induced optical activity, and $\Delta n$ is the refractive index difference from the photo-induced birefringence. The optical activity contribution to the polarization rotation $\bar{\theta}[a/b\text{ axis}]$ was extracted from the average over all probe incident angles $\varphi$. The corresponding rotary power $\rho[a/b\text{ axis}]$ is then calculated from $\bar{\theta}[a/b\text{ axis}]$ divided by the penetration depth of the excitation pulse (Fig. S4b,d). The rotary power flips sign when changing between *a*- and *b*-axis excitation, consistent with symmetry [2].

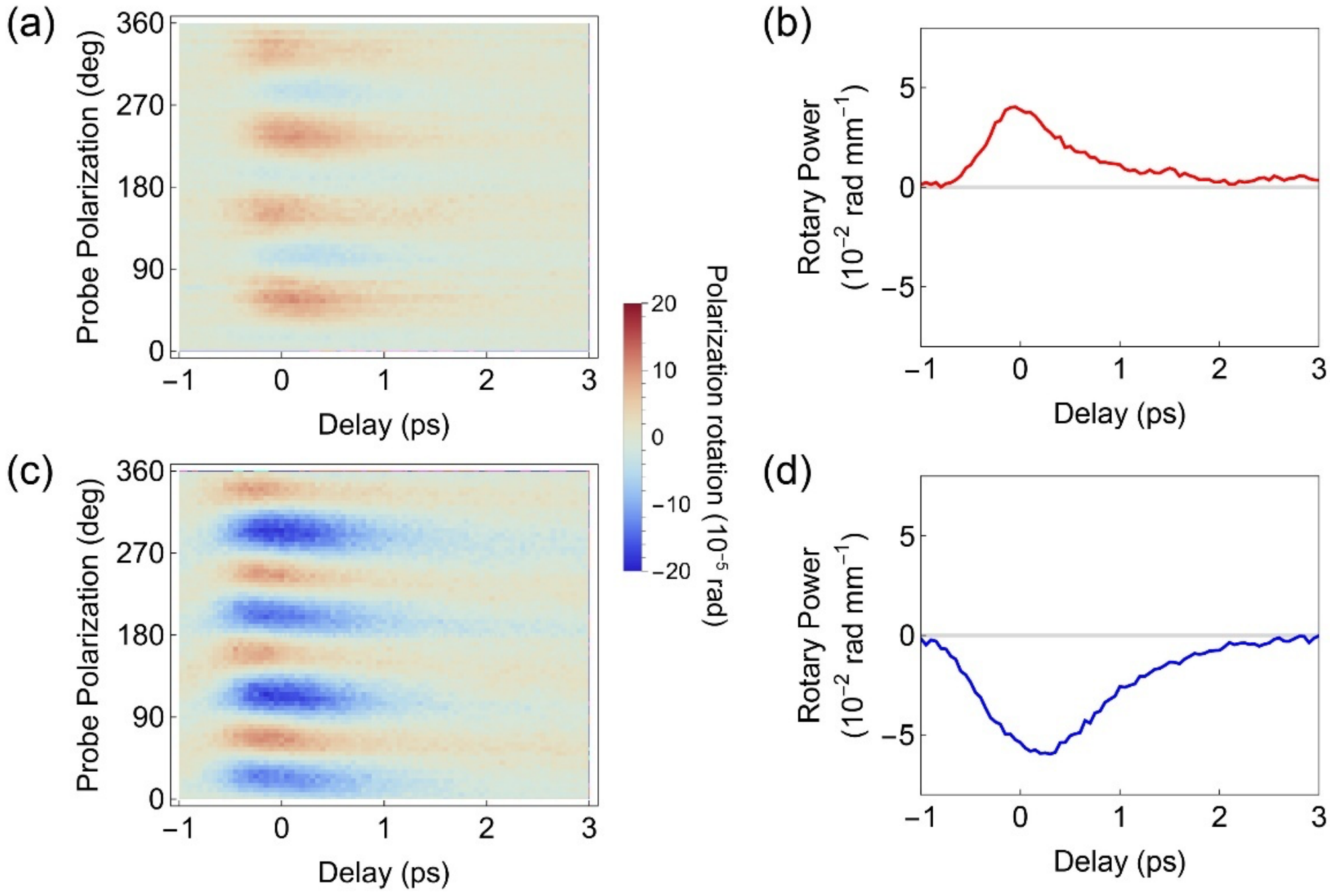

**Figure S4: Time resolved polarization rotation signal for 33.3 THz excitation. (a)** Time delay-dependent polarization rotation as a function of probe incident polarization for MIR excitation along the *a*-axis. **(b)** Time delay-dependent rotary power extracted from the data shown in (a). **(c,d)** Same as (a,b) but for MIR excitation along the *b*-axis.

We define the chirality-induced rotary power as

$$\rho = \frac{\rho[a\text{ axis}] - \rho[b\text{ axis}]}{2} \tag{S3}$$

Figure S5 shows this quantity for different excitation frequencies at comparable excitation fluences. All the time traces were fitted with a product of step function and an exponential decay

$$\rho(t) = A\left(1 + \operatorname{erf}\left(\frac{t - t_0}{s}\right)\right) e^{-t/\tau} \tag{S4}$$

with $s$ and $\tau$ being the signal rise time and the decay constant, respectively.

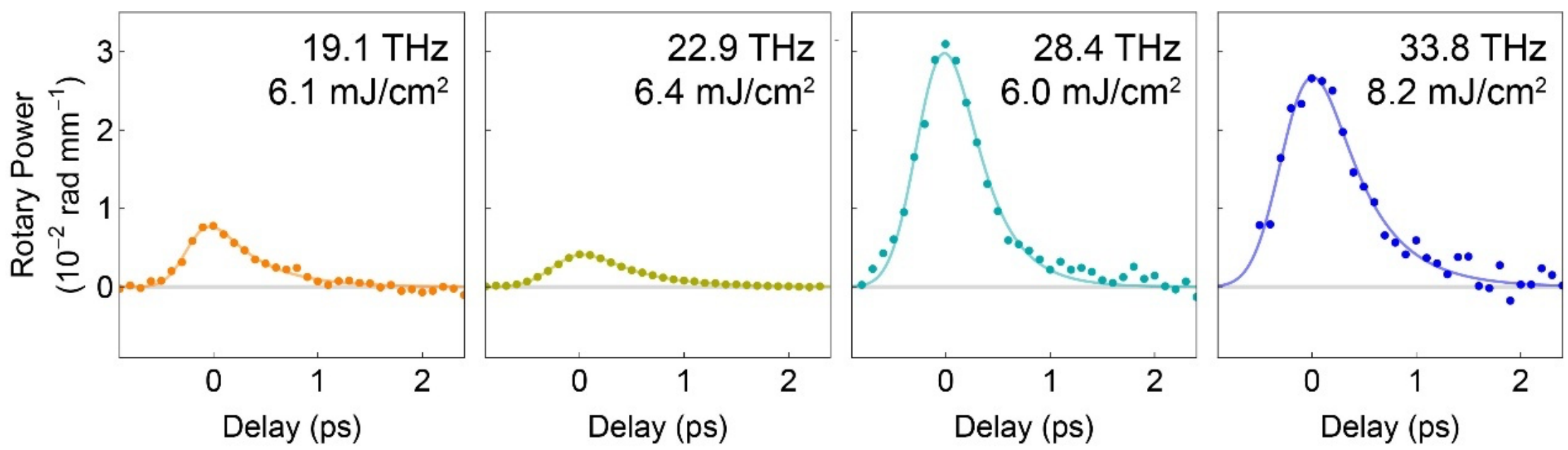

**Figure S5: Chirality-induced optical activity at different excitation frequencies.** Time-delay dependent rotary power for different excitation frequencies, measured at comparable excitation fluences.

**Section IV: Frequency dependence of the photosusceptibility**

For each excitation frequency, we measured the time delay dependent chirality-induced rotary power as a function of excitation fluence. The peak rotary power $\rho_{max}$ for each fluence was extracted by fitting Eq. (S4) to the data. This quantity $\rho_{max}$ followed a linear dependence in the excitation fluence $I$ at each the excitation frequencies (Fig. S6). The photosusceptibility was defined as the slope $\chi$ of the linear fit $\rho_{max} = \chi I$.

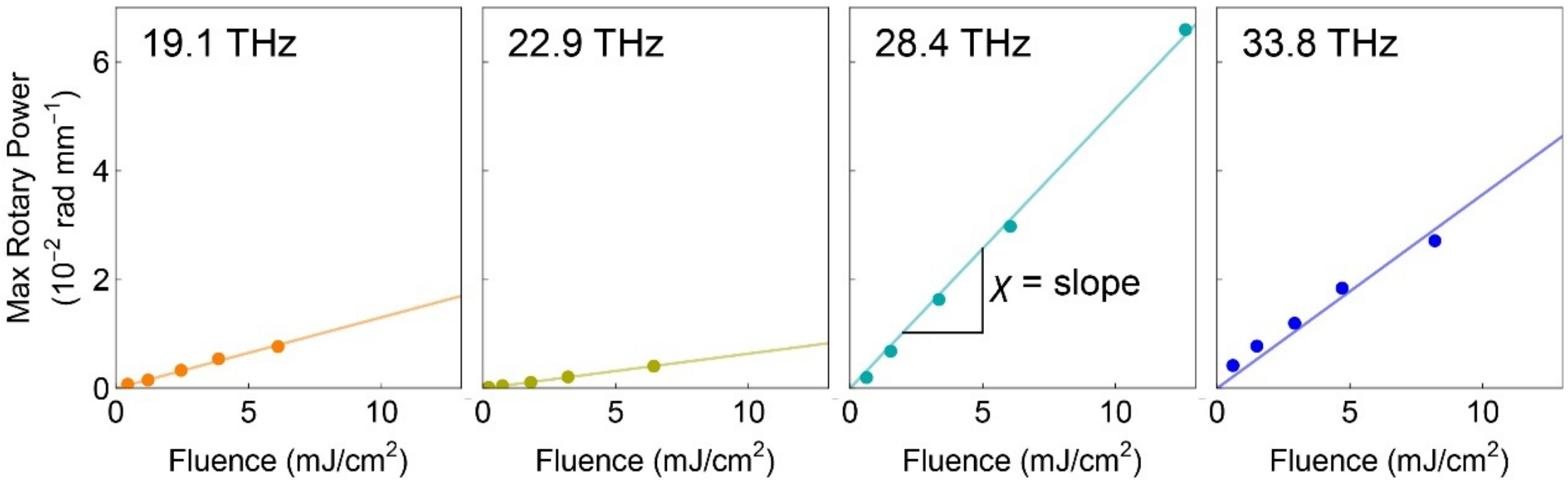

**Figure S6: Examples of the excitation fluence dependent maximum rotary power.**

**Section V: Density functional theory calculations**

We performed first-principles calculations within the framework of density functional theory (DFT) to investigate the phonon excitation spectrum, anharmonic lattice-coupling coefficients, and optical response of $BPO_4$ using the following technical and numerical settings. In general, we used the Vienna Ab initio Simulation Package (VASP) 6.6 DFT implementation [4–6], together with the Phonopy software package [7] for the phonon calculations. Our calculations employed pseudopotentials constructed within the projector augmented-wave (PAW) method [8]. Specifically, we used the default pseudopotentials for B ($2s^2$ $2p^1$), P ($3s^2$ $3p^3$), and O ($2s^2$ $2p^4$), and treated the exchange-correlation potential within the generalized gradient approximation (GGA) [9]. Lastly, as noted in Ref. [10], we employ local field corrections in the framework of the Bethe-Salpeter equation [11,12] atop the many-body GW quasiparticle framework [13,14].

After convergence checks, we used a (7 x 7 x 5) Monkhorst–Pack [10] (k)-point mesh for sampling the Brillouin zone and a plane-wave energy cutoff of 700 eV as the final numerical settings for the structural relaxations and phonon calculations. The self-consistent calculations were iterated until the change in total energy was converged to ($10^{-8}$) eV. All phonon-mode and anharmonic-coupling-constant calculations were performed in a (2 x 2 x 2) conventional cell of $BPO_4$.

The mode effective charges and dielectric response, including optical activity, were calculated using the LOPTICS settings of VASP, in addition to the approaches described in Refs. [10,16]. To achieve convergence of the optical properties, we increased the (k)-point mesh to (128 x 128 x 82) and used 1000 empty states.

The starting point of our first-principles investigation was the tetragonal cell of $BPO_4$. We first determined the DFT equilibrium volume of this cell, obtaining ($V_{cell}$ = 131.636 Å$^3$), with (a = b = 4.43 Å) and (c = 6.71 Å). The relaxed atomic equilibrium positions were found at the Wyckoff sites B (2d), P (2a), and O (8g), with oxygen coordinates (0.132, 0.258, 0.127). We then calculated the relevant phonon eigenfrequencies and mode-effective charges, which are listed in Table S2. Using the corresponding eigenvectors, we computed the coupling constants defined in Supplementary Section VI, which are listed in Table S3.

Finally, we also computed the changes in the antisymmetric parts of the dielectric tensor induced by lattice distortions along the coordinates of the four transiently displaced B-symmetry modes, the directly driven E modes and the combination of dual E modes. The resulting changes in the optical activity, are listed in Table S4.

**Table S2: Eigenfrequencies and mode effective charges obtained from DFT.**

| Mode | Frequency (THz) | $Z^*$ ($q_e/u^{1/2}$) |
|---|---|---|
| E(1) | 7.0 | 0.1 |
| E(2) | 18.1 | 0.8 |
| E(3) | 28.3 | 1.2 |
| E(4) | 32.2 | 1.4 |

| B(1) | 15.8 | 0.9 |
|---|---|---|
| B(2) | 17.7 | 0.4 |
| B(3) | 28.1 | 1.1 |
| B(4) | 31.5 | 1.5 |

**Table S3: Nonlinear coupling coefficients for $Q_E^2 Q_B$ type coupling between E- and B-modes, obtained from DFT.**

| E-mode | B-mode | $Q_E^2 Q_B$ ($meV/u^{3/2}$Å$^3$) |
|---|---|---|
| 1 | 1 | 5.6 |
| 1 | 2 | 9.1 |
| 1 | 3 | -6.8 |
| 1 | 4 | 31.6 |
| 2 | 1 | -7.5 |
| 2 | 2 | -41.3 |
| 2 | 3 | 18.6 |
| 2 | 4 | -108.6 |
| 3 | 1 | 125.0 |
| 3 | 2 | -16.2 |
| 3 | 3 | 812.1 |
| 3 | 4 | 247.4 |
| 4 | 1 | -127.7 |
| 4 | 2 | -232.1 |
| 4 | 3 | 362.5 |
| 4 | 4 | -1217.0 |

**Table S4: Nonlinear coupling coefficients for $Q_{E,1} Q_{E,2} Q_B$ type coupling between E- and B-modes, obtained from DFT.**

| E,1-mode | E,2-mode | B-mode | $Q_{E,1} Q_{E,2} Q_B$ ($meV/u^{3/2}$Å$^3$) |
|---|---|---|---|
| 1 | 2 | 1 | -33.2 |
| 1 | 2 | 2 | 15.5 |
| 1 | 2 | 3 | -61.6 |
| 1 | 2 | 4 | 64.2 |
| 1 | 3 | 1 | 37.3 |
| 1 | 3 | 2 | 66.2 |
| 1 | 3 | 3 | 156.9 |
| 1 | 3 | 4 | 60.4 |
| 1 | 4 | 1 | -93.3 |
| 1 | 4 | 2 | -22.0 |
| 1 | 4 | 3 | 77.5 |
| 1 | 4 | 4 | 36.8 |
| 2 | 3 | 1 | 82.6 |
| 2 | 3 | 2 | -55.4 |
| 2 | 3 | 3 | 359.6 |
| 2 | 3 | 4 | 147.3 |
| 2 | 4 | 1 | -81.1 |
| 2 | 4 | 2 | -93.7 |
| 2 | 4 | 3 | 214.6 |
| 2 | 4 | 4 | -497.3 |
| 3 | 4 | 1 | 257.2 |
| 3 | 4 | 2 | -18.2 |
| 3 | 4 | 3 | 1066.2 |
| 3 | 4 | 4 | 778.6 |

**Table S5: Phonon amplitude dependent changes in the rotary power connected to the light-induced chiral state, obtained from DFT.**

| Mode | $\frac{\partial^2 \rho}{\partial Q_i^2}$ ($^\circ\, mm^{-1}\, u^{-1}$Å$^{-2}$) | Mode | $\frac{\partial \rho}{\partial Q_i}$ ($^\circ\, mm^{-1}\, u^{-\frac{1}{2}}$ Å$^{-1}$) |
|---|---|---|---|
| E(1) | 1.1 | B(1) | -20.1 |
| E(2) | 0.5 | B(2) | 37.9 |
| E(3) | 7.4 | B(3) | -18.6 |
| E(4) | 9.4 | B(4) | -11.7 |

| E,1-mode | E,2-mode | $\frac{\partial^2 \rho}{\partial Q_i \partial Q_j}$ ($^\circ\, mm^{-1}\, u^{-1}$Å$^{-2}$) |
|---|---|---|
| 1 | 2 | 6.2 |
| 1 | 3 | 7.0 |
| 1 | 4 | 11.8 |
| 2 | 3 | 7.6 |
| 2 | 4 | -9.4 |
| 3 | 4 | -7.2 |

**Section VI: Simulation of phonon-polariton propagation and nonlinear dynamics**

We simulated the space- and time-dependent electric field and phonon mode dynamics induced by the MIR excitation pulse, using the method of lines with discrete spatial coordinate.

Inside the $BPO_4$ sample ($0 < z < 150\ \mu m$, with $z$ denoting the spatial coordinate along the *c*-axis propagation direction of the plane wave), the E-symmetry modes carry a finite dipole in the *ab*-plane, and couple linearly to the electric field, forming the phonon polariton. The coupled differential equation for the E-symmetry mode coordinates $Q_{E,i}$ and the electric field $E$ can be written as [13]

$$\frac{\partial^2 Q_{E,i}}{\partial t^2} + \gamma_{E,i}\frac{\partial Q_{E,i}}{\partial t} + \omega_{E,i}^2 Q_{E,i} = Z_{E,i}^* E \tag{S5}$$

$$\frac{\partial^2 E}{\partial z^2} = \frac{1}{c^2}\frac{\partial^2}{\partial t^2}\left(\varepsilon E + \sum_i \beta_{E,i} Q_{E,i}\right) \tag{S6}$$

where $c$ is the speed of light, $\varepsilon$ is the dielectric constant, $\gamma_{E,i}$ are damping coefficients, $\omega_{E,i}$ are phonon frequencies, $Z_{E,i}^*$ and $\beta_{E,i}$ are the effective charge and dipole of phonons, respectively. The parameters related to E-modes are calculated from the FTIR results in Table S1.

Outside the sample ($z > 150\ \mu m$ or $z < 0$), the electric field $E$ propagates according to the wave equation in vacuum

$$\frac{\partial^2 E}{\partial z^2} = \frac{1}{c^2}\frac{\partial^2 E}{\partial t^2} \tag{S7}$$

The initial condition for the simulation is set by a Gaussian electromagnetic field outside the sample that propagates along the $+z$ direction

$$E(z, t = 0) = E_0 e^{-\frac{(z-z_0)^2}{(c\tau_d)^2}} \cos\left(\frac{\omega_0 z}{c}\right) \tag{S8}$$

where $\tau_d$ is the duration of the excitation pulse, and $\omega_0$ is the central frequency. The initial central coordinate $z_0 < -3c\tau_d$ ensures that the pulse is outside of the sample at $t = 0$. In addition, we take $2\sqrt{\ln 2}\,\tau_d = 600\ fs$ to match the simulations to the experimental conditions. Examples of the simulated space- and time-dependent electric field $E$ and phonon coordinate $Q_{E,i}$ for excitations on resonance and off resonance with the E(4) mode are shown in Fig. S7a,d and Fig. S7b,e, respectively.

Next, we consider the nonlinear coupling of the driven modes $Q_{E,i}$ to the B-symmetry phonons $Q_{B,k}$. The equation of motion for these modes, according to the interaction potential given by Eq. (2) in the main text, is

$$\frac{\partial^2 Q_{B,k}}{\partial t^2} + \gamma_{B,k}\frac{\partial Q_{B,k}}{\partial t} + \omega_{B,k}^2 Q_{B,k} = \pm\left[\sum_i \alpha_{i,k} Q_{E,i}^2 + \sum_i \delta_{i,k} E Q_{E,i} + \zeta_k E^2 + \sum_{i<j} \kappa_{ij,k} Q_{E,i} Q_{E,j}\right] \tag{S9}$$

where $\gamma_{B,k}$ are damping coefficients, $\omega_{B,k}$ are phonon frequencies, and $\alpha_{i,k}$, $\delta_{i,k}$, $\zeta_k$, and $\kappa_{ij,k}$ are the two-phonon, infrared resonant Raman, electronic Raman, and three-phonon coupling coefficients, respectively. The nonlinear coupling coefficients are taken from first-principle calculations (Tables S3 and S4). The sign on the right-hand side of Eq. (S9) is determined by the polarization of the excitation pulses (along $a$ or $b$), as determined by the crystal symmetry. Note that we do not consider linear coupling of the drive electric field to the B-symmetry modes, because their dipoles are oriented parallel to the $c$-axis. We show one example of the simulated $Q_{B,k}$ dynamics in Fig. S7c and S7f for the same excitations shown in Fig. S7a,b and Fig. S7d,e, respectively.

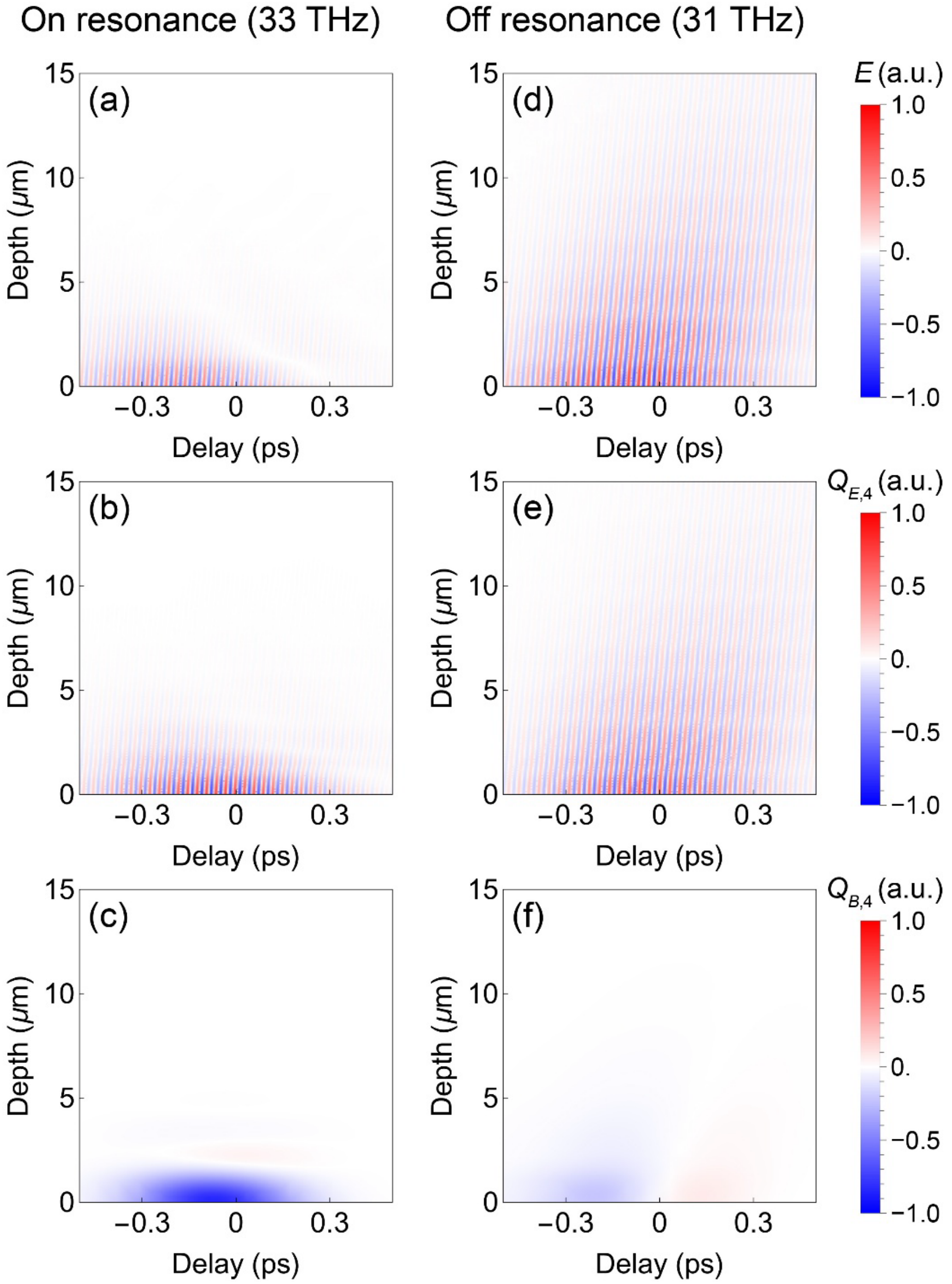

**Figure S7: Example results of time- and space-dependent phonon-polariton simulations. (a)** Temporal and spatial evolution of the electric field $E$ for 33-THz excitation. **(b)** Temporal and spatial evolution of the E-symmetry mode $Q_{E,4}$ for the same on-resonance 33-THz excitation. **(c)** Temporal and spatial evolution of the B-symmetry mode $Q_{B,4}$, nonlinearly coupled to the electric field shown in panel (a) and the E(4) mode shown in panel (b). The coupling coefficients are taken

from first-principle calculations. The result is low pass filtered along the time axis, showing only the rectified component. **(d-f)** Same as (a-c) but for 31-THz excitation, i.e. below the resonance of the E(4) mode $Q_{E,4}$.

The time- and space-dependent E- and B-symmetry mode displacements $Q_{E,i}$ and $Q_{B,k}$, allowed us to quantify the rotary power $\rho(z,t)$ at each position in the sample as the linear combination

$$\rho(z,t) = \sum_k \left(\frac{\partial \rho}{\partial Q_{B,k}}\right) Q_{B,k}(z,t) + \frac{1}{2}\sum_{i,j}\left(\frac{\partial^2 \rho}{\partial Q_{E,i}\partial Q_{E,j}}\right) Q_{E,i}(z,t) Q_{E,j}(z,t) \qquad \text{(S10)}$$

Here, the constants $g_{B,k} = \partial\rho/\partial Q_{B,k}$ for B-symmetry modes and $g_{E,ij} = \partial^2\rho/\partial Q_{E,i}\partial Q_{E,j}/2$ for E-symmetry modes are taken from the first-principle calculations (Table S5).

In the time-space diagrams shown in Fig. S7, NIR probe pulses propagate along lines with the slope of $c/n_g$ (with $n_g$ being the group index at 800 nm). The overall polarization rotation experienced by the probe light can be calculated by integration along the lines of propagation [2], as

$$\theta(t) = \int_0^{150\,\mu m} \frac{1}{\sqrt{\pi}\tau_p}\int_{-\infty}^{\infty} e^{-\frac{t'^2}{\tau_p^2}} \rho\left(z', t + \frac{z'}{c/n_g} + t'\right) dt'\, dz' \qquad \text{(S11)}$$

Here, the finite probe pulse duration $\tau_p$ is considered with the integration over $dt'$. The integration of $dz'$ sums the contributions from all the layers inside the sample. The time-resolved polarization rotation signals simulated from this integration, which corresponds to the $Q_{E,i}$ displacements in Fig. S7b,e and $Q_{B,k}$ displacements in Fig. S7c,f, respectively, are shown in Fig. S8 and Fig. S9.

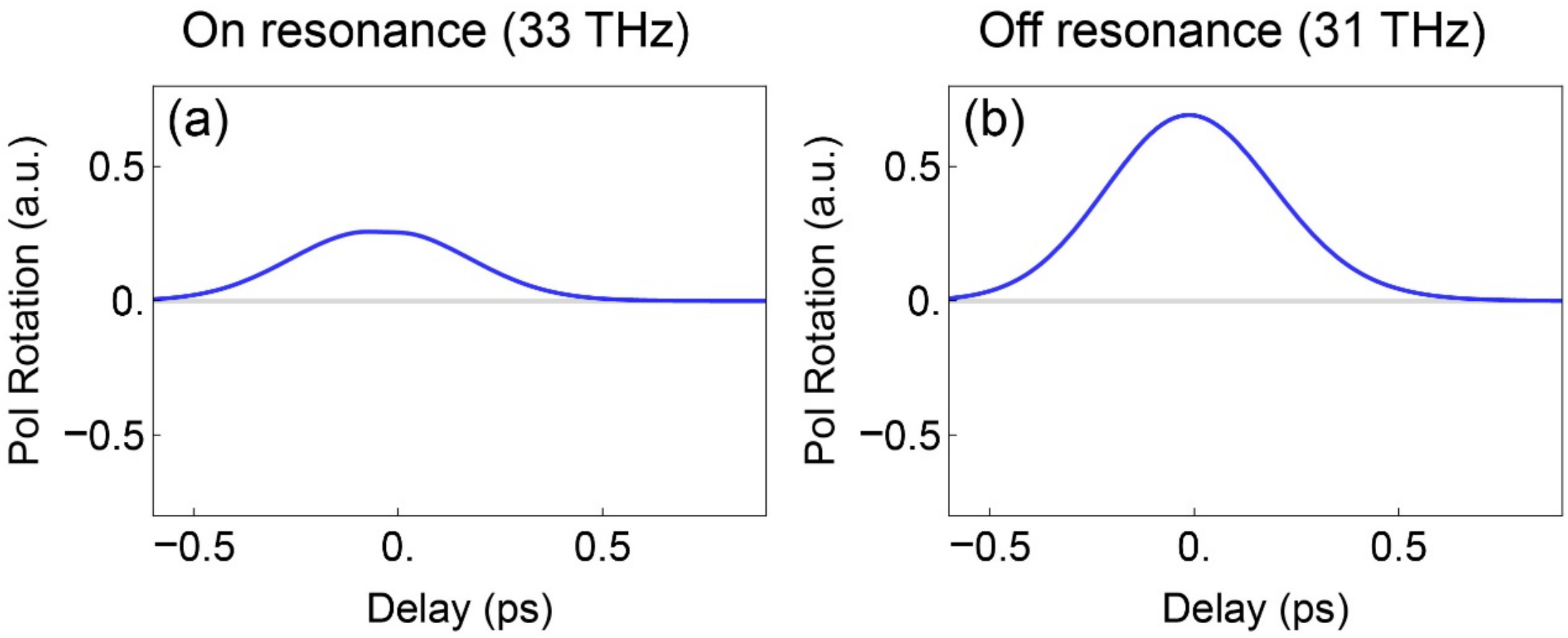

**Figure S8: Example results of the simulated time-resolved polarization rotation from the E-modes, considering $g_{E,ii}Q_{E,i}^2$ and $g_{E,ij}Q_{E,i}Q_{E,j}$ contributions. (a)** and **(b)** are integrated from Fig. S7b,e, respectively.

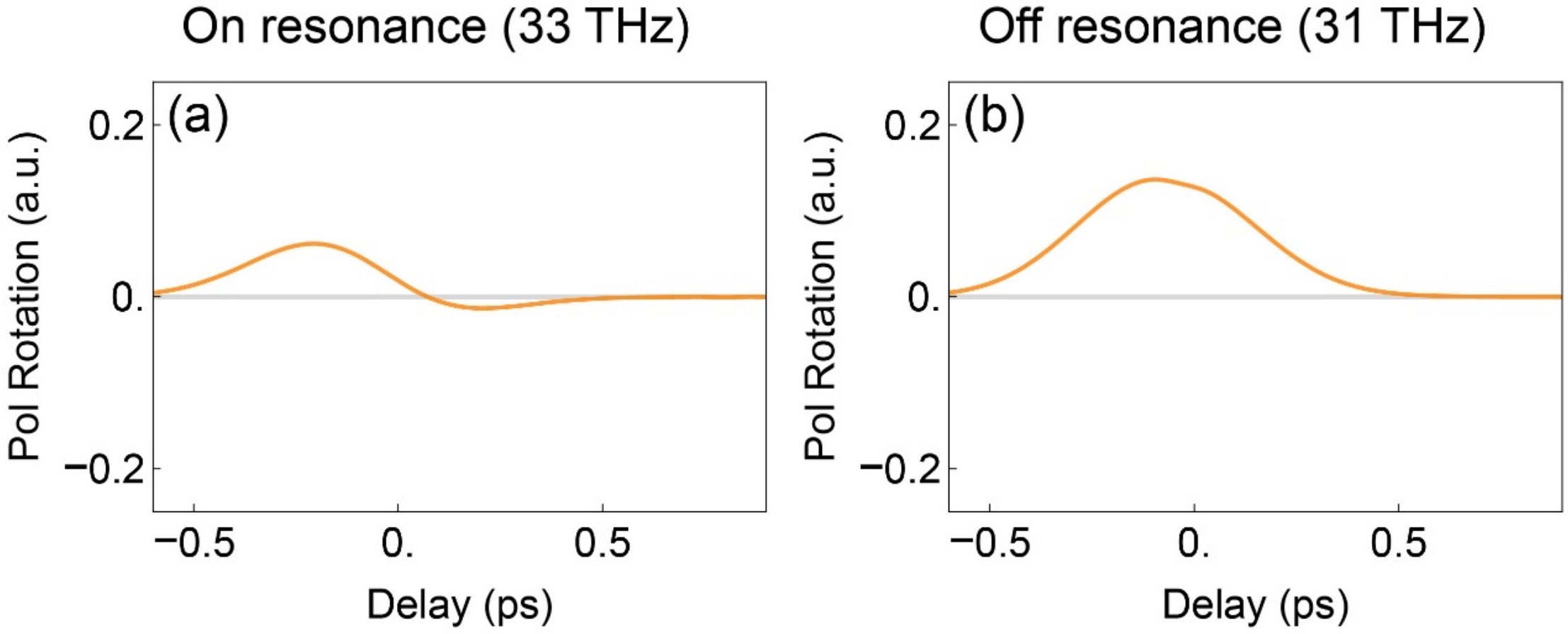

**Figure S9: Example results of the simulated time-resolved polarization rotation from the B-modes, considering $g_{B,k}Q_{B,k}$ contributions. (a)** and **(b)** are integrated from Fig. S7c,f, respectively.

We simulated the separate contributions to the polarization rotation from the $g_{B,k}Q_{B,k}$ terms related to the nonlinearly coupled to the B-modes $Q_{B,k}$, as well as the $g_{E,ii}Q_{E,i}^2$ and $g_{E,ij}Q_{E,i}Q_{E,j}$ terms, related to the optically driven E-modes. Because the differential equations ensure a linear dependence of the chirality-induced polarization rotation on the excitation fluence, we performed all the simulations at one fixed incident electric field, and extracted the photosusceptibility by division through the corresponding excitation fluence (Fig. 4c in the main text).

**Section VII: Fit to the experimental frequency-dependent photosusceptibility**

The simulations of the chirality-induced rotary power were based on the nonlinear coupling coefficients and gyration coefficients derived from first-principle calculations. However, quantitative deviations from experiment are unavoidable because the first-principle approach necessarily involves approximations in the exchange-correlation functional and idealized structural conditions. In addition, the experiments may be influenced by defects and sample inhomogeneity. Therefore, a reasonable strategy is to permit the parameters, especially the optical gyration coefficients, to vary around their DFT-predicted values, yielding a fit based on the experimental data that remains guided by the first-principle calculations.

Because the measured optical activity contains inseparable contributions from the rectified displacements of the different B-modes and directly from the self-rectified and mixed E-modes, the data do not uniquely constrain all components of the gyration coefficients. Different pathways with the same excitation frequency dependence can be grouped together, which leads to

$$\rho = \pm\left[\sum_i A_i \langle Q_{E,i}^2 \rangle + \sum_{i<j} K_{ij} \langle Q_{E,i} Q_{E,j} \rangle\right] \tag{S12}$$

where $\langle \dots \rangle$ is the time average, and $A_i, K_{ij}$ are the weighted summation of the gyration coefficients with the same frequency dependence over all pathways

$$A_i = \sum_k \frac{\alpha_{i,k}}{\omega_{B,k}^2} \frac{\partial \rho}{\partial Q_{B,k}} + \frac{1}{2} \frac{\partial^2 \rho}{\partial Q_{E,i}^2}, K_{ij} = \sum_k \frac{\kappa_{ij,k}}{\omega_{B,k}^2} \frac{\partial \rho}{\partial Q_{B,k}} + \frac{\partial^2 \rho}{\partial Q_{E,i} \partial Q_{E,j}} \tag{S13}$$

In our fitting procedure, with the goal of matching the frequency-dependent photosusceptibility with the rotary power given by Eqs. (S12) and (S13), the free parameters are the coefficients $A_i$ and $K_{ij}$. Because our experiments cover the resonance frequencies of E(2), E(3), and E(4) modes, we introduce six fitting parameters in total, including three coefficients $A_i$ (with $i = 2,3,4$) and three coefficients $K_{ij}$ (with $ij = 23,24,34$). Initial values of these coefficients are calculated from the first-principle data summarized in Tables S2–S5.

The result of the numerical fit is plotted together with the experimental data in Fig. 5 of the main text. Table S6 shows the initial values and the converged fitted values of the ratios $|K_{ij}/A_i|$ and $|K_{ij}/A_j|$ for the different E-symmetry modes, which characterize the relative size of the terms proportional to $Q_{E,i}^2$ and $Q_{E,i}Q_{E,j}$. The similar size of $K_{ij}$ and $A_i$ within about one order of magnitude is consistent with the fact that the two-phonon and three-phonon mechanisms share the same order of lattice anharmonicity, and signifies the importance of both mechanisms.

**Table S6: Ratios $|K_{ij}/A_i|$ and $|K_{ij}/A_j|$ for different E-symmetry mode pairs $i, j$ from the fit to the experimental results.**

| Quantity | DFT result | Fitting result |
|---|---|---|
| $\lvert K_{34}/A_3 \rvert$ | 0.18 | 2.58 |
| $\lvert K_{34}/A_4 \rvert$ | 0.25 | 1.78 |
| $\lvert K_{24}/A_2 \rvert$ | 5.15 | 14.5 |
| $\lvert K_{24}/A_4 \rvert$ | 0.51 | 1.10 |
| $\lvert K_{23}/A_2 \rvert$ | 9.88 | 8.25 |
| $\lvert K_{23}/A_3 \rvert$ | 0.68 | 0.90 |